\documentclass[12pt,preprint]{aastex}
\usepackage{emulateapj5}

\begin{document}

\title{Non--equilibrium CO chemistry in the solar atmosphere}

\author{A. Asensio Ramos, J. Trujillo Bueno\altaffilmark{1}}
\affil{Instituto de Astrof\'{\i}sica de Canarias, 38200 La Laguna, Tenerife, Spain}
\altaffiltext{1}{Consejo Superior de Investigaciones Cient\'{\i}ficas (Spain).}
\email{aasensio@ll.iac.es, jtb@ll.iac.es}
\author{M. Carlsson}
\affil{Institute for Theoretical Astrophysics, P.O. Box 1029, Blindern, N-0135, Oslo, Norway}
\email{Mats.Carlsson@astro.uio.no}
\and
\author{J. Cernicharo\altaffilmark{1}}
\affil{Departamento de Astrof\'{\i}sica Molecular e Infrarroja. 
Instituto de Estructura de la Materia\\ Serrano 123, E-28006 Madrid, Spain}
\email{cerni@astro.iem.csic.es}


\begin{abstract}
 
Investigating the reliability of the assumption of instantaneous chemical equilibrium
(ICE) for calculating the CO number density in the 
solar atmosphere is of crucial importance
for the resolution of the long-standing controversy over the existence of 
`cool clouds' in the chromosphere, and for determining whether
the cool gas owes its existence to CO radiative cooling or to a hydrodynamical
process. Here we report the first results
of such an investigation in which we have carried out time-dependent
gas-phase chemistry calculations in radiation hydrodynamical
simulations of solar chromospheric dynamics. We show that while 
the ICE approximation turns out to be suitable
for modeling the observed infrared CO lines at the solar disk center,
it may substantially overestimate the `heights of formation' of strong
CO lines synthesized close to the edge of the solar disk, especially
concerning vigorous dynamic cases resulting from relatively strong photospheric 
disturbances. This happens because during the
cool phases of the hydrodynamical simulations the CO number density
in the outer atmospheric regions is smaller than what is stipulated by the ICE approximation,
resulting in decreased CO opacity in the solar chromosphere. As a result,
the cool CO-bearing gas which produces the observed molecular lines  
must be located at atmospheric heights not greater than 700 km, approximately.
We conclude that taking into account the non-equilibrium chemistry
improves the agreement with the available on-disk and off-limb observations, 
but that the hydrodynamical simulation model has to be even cooler than
anticipated by the ICE approximation, and this has to be the case at the `new' (i.e. deeper)
formation regions of the rovibrational CO lines.
\end{abstract}


\section{Introduction}
\label{introduction}

Thirty years ago Noyes \& Hall (1972) inferred
very low brightness temperatures (T$\,{\approx}\,3700$ K)
from their discovery of strong rovibrational CO lines  
at 4.7 $\mu{\rm m}$ observed close to the edge of the solar disk. 
Several years later, after such surprising observational results
had been confirmed by Ayres \& Testerman (1981),
it was suggested that the low chromosphere might not be
hot at all but could instead be permeated by CO-cooled `clouds'
at altitudes between 500 and 1000 kilometers 
above continuum optical depth unity (Ayres 1981). 
This led to controversy because other (UV and submillimeter) diagnostics
had suggested the existence of a uniformly hot chromosphere
with a minimum temperature of about 4400 K near 500 km
and a temperature rise above this temperature-minimum region.
The controversy over the existence of cool gas in the low
chromosphere continues today (see Kalkofen 2001, Ayres 2002, Avrett 2003), 
after the publication of an abundance of literature on the subject with theoretical investigations
concluding that the CO lines have LTE source functions
(Ayres \& Wiedemann 1989; Uitenbroek 2000a) and with new observations
showing off-limb CO emission protruding hundreds of kilometers into
the chromosphere (Solanki, Livingston \& Ayres 1994; Clark et al. 1995), but also
with the discovery that far-UV chromospheric lines observed on the solar disk
always remain in emission at all positions and times (Carlsson, Judge \& Wilhem 1997).

Over the last few years, it has become increasingly evident that
the next crucial step towards a better understanding
of the enigmatic thermal structure of the solar chromosphere
is to carefully investigate the reliability of the
assumption of instantaneous chemical equilibrium (ICE), which is
currently used for calculating the molecular number densities
in stellar atmospheres (see, in particular, Uitenbroek 2000a,b; 
see also Ayres \& Rabin 1996, Avrett et al. 1996,
Asensio Ramos \& Trujillo Bueno 2003, Avrett 2003). 
Actually, both observations and simulations 
indicate that the solar chromosphere is a highly inhomogeneous and dynamic
region of low density plasma whose thermal, dynamic and magnetic
properies we need to decipher for unlocking new discoveries.
If the ICE approximation turns out to be adequate for modeling
the strongest CO lines close to the edge of the solar disk, 
then the available CO observations
would really be indicating the existence of cool gas in the solar chromosphere. 
Otherwise, a natural resolution of the current chromospheric temperature
discrepancy could perhaps emerge
if the CO concentration turns out to be sufficiently lower 
than would be expected on the basis of chemical equilibrium.
This Letter addresses this challenging issue by showing
the first results of a detailed investigation on the non-equilibrium CO chemistry
in the solar atmosphere. 

\section{Formulation of the problem}
\label{chemical_evolution}
Our strategy consists in performing chemical evolution calculations in  
the radiation hydrodynamical simulations 
of solar chromospheric dynamics described by Carlsson \& Stein (1997, 2002),
that do not include CO cooling in the energy equation.
Therefore, at each time step of the hydrodynamical simulation,
we have fixed the ensuing thermodynamic conditions
and calculated the corresponding CO number density by following the
chemical evolution starting from the molecular concentrations
of the previous time step.

The chemical evolution (CE) problem consists in calculating
the molecular number densities at each time step of the hydrodynamical simulation
by taking into account all relevant chemical reactions with their ensuing rates. 
In order to solve
this problem, one has to consider the following evolution equation for 
each chemical species `$i$' included in the model: 

\begin{eqnarray}
\label{chem_sys}
& \frac{dn_i}{dt} = \sum_A \sum_B \sum_C k_{ABC} n_A n_B n_C + \sum_A \sum_B k_{AB} n_A n_B + \nonumber \\
& + \sum_A k_{A} n_A - \sum_A \sum_B k_{ABi} n_A n_B n_i - \nonumber \\
& - \sum_A k_{Ai} n_A n_i - k_i n_i,
\end{eqnarray}
where \emph{three-body} reactions (first and fourth terms), \emph{two-body} 
reactions (second and fifth terms) and \emph{one-body}
reactions (third and
sixth terms) have been taken into account. 
When all the reaction rates for three-body ($k_{ABC}$), 
two-body ($k_{AB}$) and one-body 
reactions ($k_{A}$) are given, this
set of equations for all the species included in the model form a set of first order 
ordinary differential equations which has to be solved numerically. Due to the \emph{stiffness} 
of the system of equations (produced
by the enormous differences in the abundances and their temporal variability), 
an implicit scheme has to
be used. Two key ingredients have to be taken into account: 
the number of chemical species and the 
reaction rates for all possible
reactions. After a careful investigation, we found that at least the following 
set of 13 species, which includes the most abundant diatomic molecules, is required
to achieve a suitable description of the relevant chemical processes: 
H, C, O, N, He, CH, CO, H$_2$, OH, NH, N$_2$, NO and CN. We have verified that the inclusion of 
ionic species does not significantly affect
the CO concentration for the typical atmospheric conditions 
encountered at heights $h{\le}1000$ km, 
although they should be ideally taken into account for very strong shocks capable
of producing sizable changes in the degree of ionization. 
Concerning the reactions rates, we have used a reaction database created for the study
of combustion mechanisms (Konnov 2000) which seems to be appropriate for the physical conditions 
in the solar atmospheric plasma. Other databases like UMIST (Bennet 1988) have been 
used for the study of the 
interstellar chemistry, but they suffer from the lack of certain important reactions which can 
efficiently take place in the Sun's atmosphere 
(e.g. the catalytic three-body formation of molecular hydrogen: H+H+He$\rightarrow$H$_2$+He).
We have also investigated the possible influence of CO photodissociations, which
are one-body reactions. They are mainly produced by 
discrete photon absorptions at wavelengths between the Lyman cutoff (912 \AA) and
the dissociation threshold (1120 \AA). By using the photodissociation rates of 
Mamon, Glassgold \& Huggins (1988), we have verified that the contribution of
photodissociation processes to the total CO concentration is negligible for 
the radiation field in the solar atmosphere. 

\section{Results and discussion}
\label{results}

We used two time-series of snapshots from the above-mentioned
radiation hydrodynamical simulations, each one lasting about 3600 seconds
and showing the upward propagation of acoustic wave trains growing
in amplitude with height until eventually producing shocks.
The first one corresponds to a relatively strong photospheric 
disturbance showing well-developed cool phases and pronounced hot zones
at chromospheric heights (see Carlsson \& Stein 1997).
The strongest dynamic cycle of this simulation
produces a peak-to-peak line-core brightness temperature of 1000 K, concerning
the strong 3-2 R14 CO-line synthesized at disk center with the ICE approximation.
However, the brightness temperature variations in
most of the cycles of this simulation are 400 K approximately, which is
similar to the observed values found by Uitenbroek, Rabin \& Noyes (1994)
under excellent seeing conditions,
but larger than those inferred from the temperature histograms in Ayres \& Rabin (1996).  
We shall refer to this as the strongly dynamic 
case.\footnote{This is similar to that used by Uitenbroek (2000a) and Ayres (2002)
in their ICE modeling, although
they considered only a segment of 190~s corresponding to the most
dynamic cycle of the full simulation.}
The second simulation corresponds to a much less intense
photospheric disturbance (see Carlsson \& Stein 2002), 
which for strong CO lines synthesized at disk center 
produces a peak-to-peak line-core brightness temperature
fluctuation which is always smaller than 400 K. 
We shall refer to this as the weakly dynamic case.  

Starting from the molecular concentration 
given by the ICE approximation, 
we have followed the chemical evolution 
in order to obtain the temporal variation of the 
CO number density ($N_{\mathrm{CO}}$) 
at each height in the simulated solar atmosphere.
As expected, the CO concentration is anticorrelated
with the local temperature variations, yielding relatively low 
$N_{\mathrm{CO}}$ values in the hot phases and relatively high $N_{\mathrm{CO}}$ values
in the cool phases. However, the amplitude of the local $N_{\mathrm{CO}}$ fluctuation
is smaller than that given by the ICE approximation. 
In relatively low density regions characteristic 
of the outer atmospheric layers (e.g. around $h=1000$ km)
the ICE approximation does a fairly good job during the hot phases,
but it overestimates the CO number density during the cool phases.
In contrast, in relatively high density regions
characteristic of photospheric layers (e.g. around
$h=400$ km) the ICE approximation underestimates the $N_{\mathrm{CO}}$
values during the hot phases. Thus, the CO abundance does
not react instantaneously to the changes in the temperature
because of the finite reaction rates. 
If one carries out a linear analysis by 
introducing a {\em small} temperature perturbation
in a medium of given density and temperature and calculates the relaxation
time needed to recover the original CO concentration one finds that, for
a given density, the relaxation time is the larger the cooler the 
medium where the temperature perturbation is introduced. Similarly,
for a given temperature of the unperturbed medium, the relaxation
time increases rapidly with decreasing density. Short relaxation
times are typical of high-temperature and high-density 
media (e.g. $t_{\rm relax}\,{\approx}\,0.006$ s for $n_{\rm H}=10^{16}\,{\rm cm}^{-3}$
and $T=6000$ K), while
long relaxation times are characteristic of low-temperature and low-density
situations (e.g. $t_{\rm relax}\,{\approx}\,400$ s for $n_{\rm H}=10^{14}\,{\rm cm}^{-3}$
and $T=4000$ K). One would find a broad range of relaxation times
at a fixed height (e.g. $h=700$ km) in the dynamic atmosphere 
that results from the above-mentioned hydrodynamical simulations,
simply because the existing rarefactions, compressions and temperature
fluctuations are continually changing the atmospheric conditions. 
Obviously, the situation is highly non-linear and the relaxation time
concept, although useful, loses its meaning in the real Sun. Therefore, any 
firm conclusion needs to be achieved via detailed numerical simulations. 

\begin{figure}
\plottwo{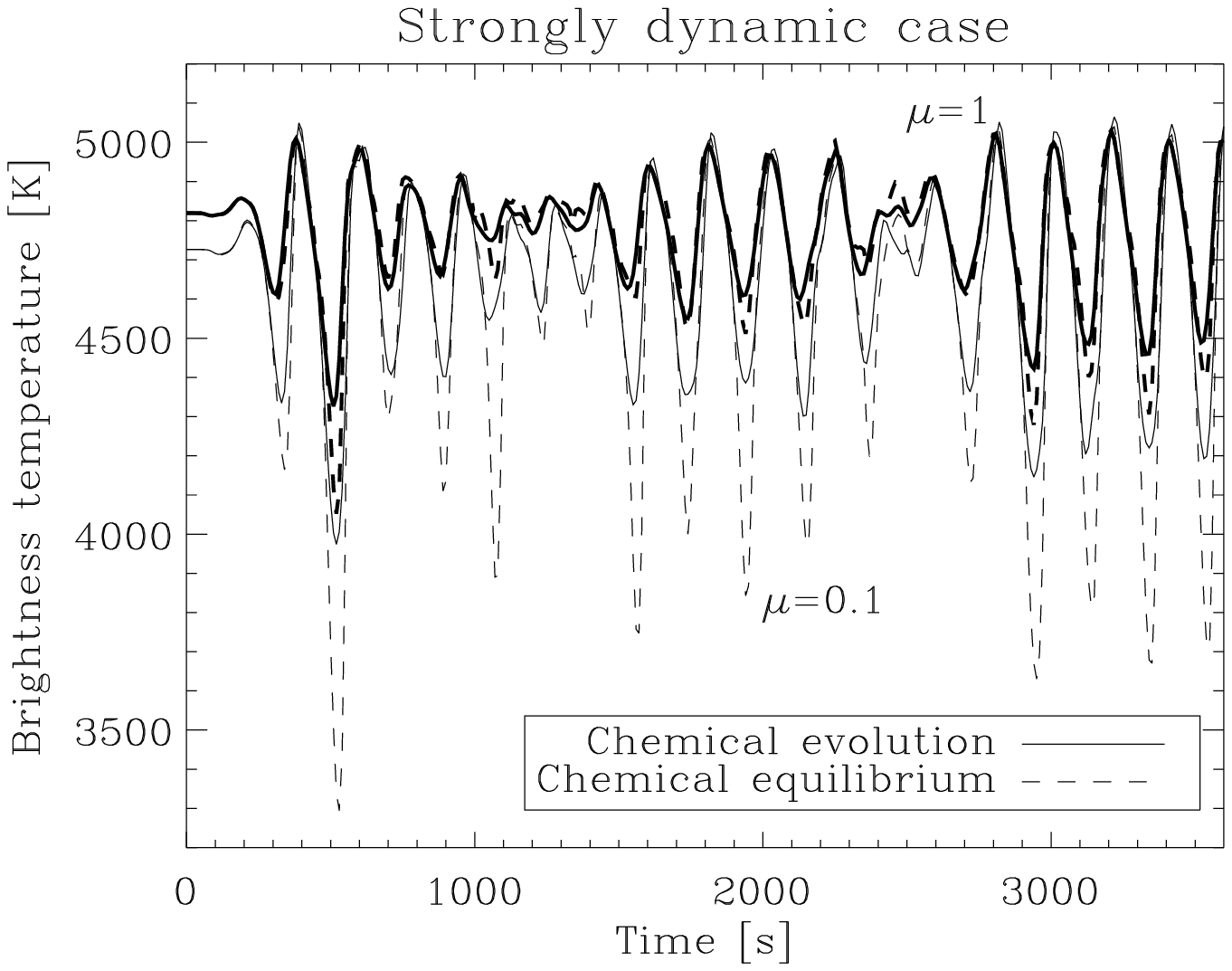}{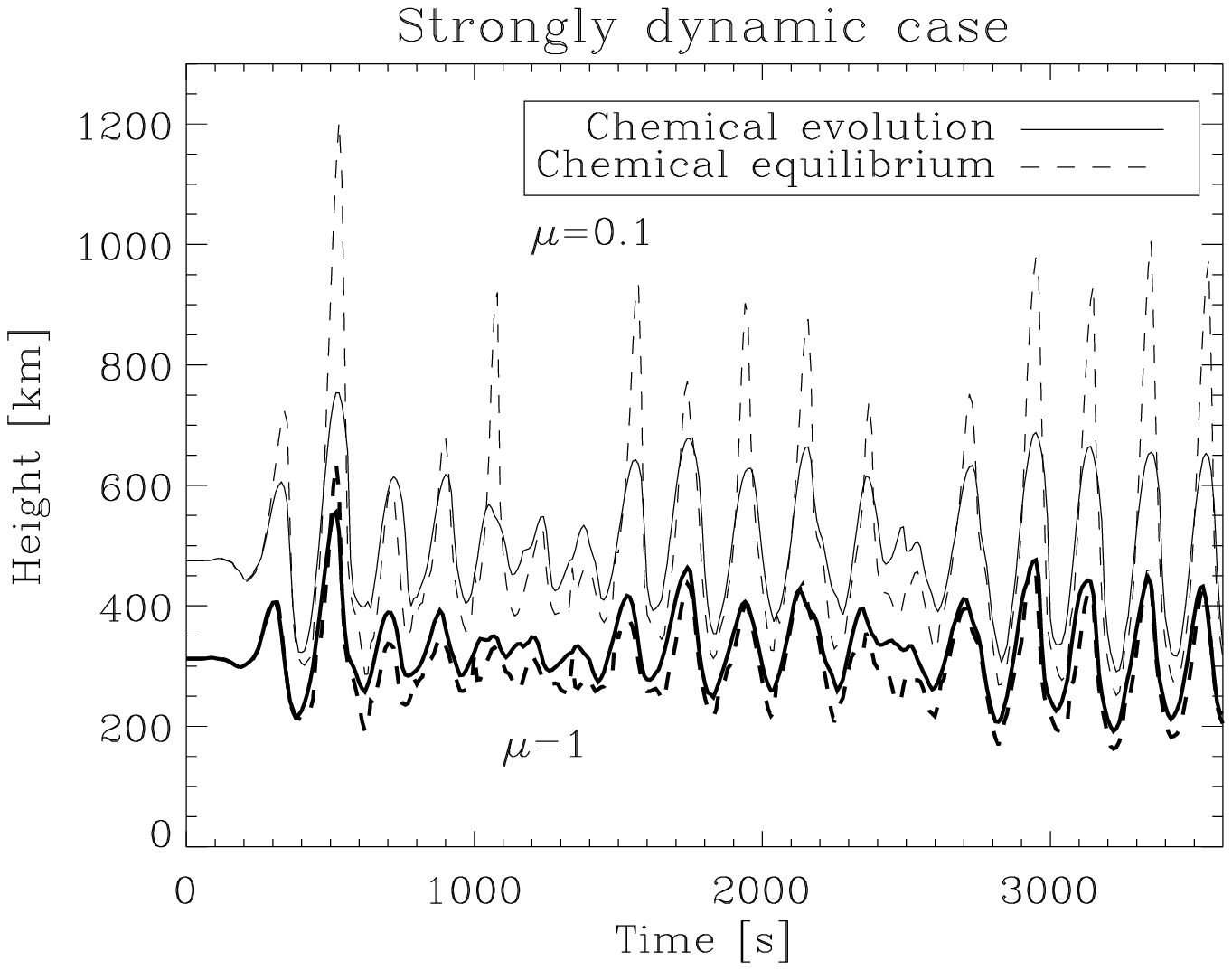}
\caption{Temporal variation of the brightness temperature in the core of the strong 
3-2 R14 line (\emph{left panel}) and of the height
of line-core optical depth unity (\emph{right panel}) for the 
strongly dynamic case and for two observing angles:
disk center ($\mu=1$, heavy lines) and close to the solar limb ($\mu=0.1$, light lines),
where $\mu={\rm cos}{\theta}$ (with $\theta$ the angle between
the solar radius vector and the line of sight).\label{fig1}}
\end{figure}

Fig. \ref{fig1} refers to the strongly dynamic case.
The left panel shows the temporal variation in the
brightness temperatures of the line-core emergent intensities
at $\mu=1$ and $\mu=0.1$ in the strong 3-2 R14 CO-line.
The right panel gives the ensuing fluctuations of the atmospheric
height where the line-core optical depth is unity, which we use
as an indicator of the `representative height'
where the CO line-core radiation originates. As seen in the
figure, the ICE approximation does a sufficiently good job concerning
the synthesis of the emergent CO spectrum at the
solar disk center ($\mu=1$), but it largely underestimates
the line-core emergent intensities at $\mu=0.1$ 
during the cool phases of the hydrodynamical simulations, 
producing brightness
temperatures that are typically 500 K {\em lower} than
those computed with the non-equilibrium CO concentration. 
Clearly, this is because during the cool phases 
the ICE approximation overestimates the `heights of line formation' 
by about 300 km, concerning the synthesis of strong
CO lines at $\mu=0.1$ in the strongly dynamic case.
This happens because during such cool episodes the CO number density
in the outer atmospheric regions is smaller than what is stipulated by the ICE approximation,
resulting in decreased CO opacity in the solar chromosphere.
Interestingly, in the weakly dynamic case
which has smaller kinetic temperature fluctuations 
(but still larger than the fluctuations of observed brightness temperatures!) 
the ICE approximation does a much better job even at $\mu=0.1$.

The reader may find surprising our conclusion that
the ICE approximation is suitable for modeling
the CO spectrum at the solar disk center, given that
Uitenbroek (2000b) found that the
spatially averaged line cores of weak CO lines synthesized
in the three-dimensional (3D) hydrodynamical model of Stein \& Nordlund (1989)
are overly dark compared to the observed ATMOS spectrum described by Farmer \& Norton (1989).
As we shall show in a forthcoming publication
(Asensio Ramos \& Trujillo Bueno; in preparation), this is because
such a 3D hydrodynamical model of the solar photosphere
is too cold in the CO line forming region. In fact, our
ICE synthesis of the 7-6 R68 CO line in the improved 3D 
hydrodynamical model of Asplund et al. (2000)
shows a notable agreement with the observed ATMOS spectrum, which constitutes
an additional indication of the realism of the most recent 
3D hydrodynamical simulations of solar surface convection
(see Shchukina \& Trujillo Bueno 2001 concerning 
the iron spectrum in such a 3D hydrodynamical model).     

\begin{figure}
\epsscale{0.5}
\plotone{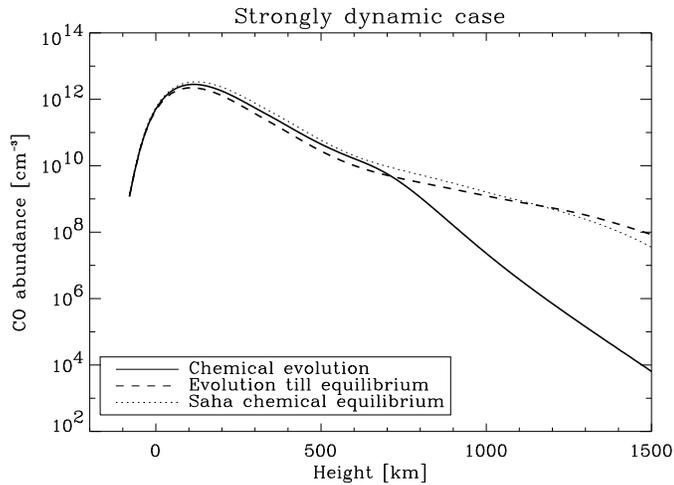}
\caption{Solid line: the height variation of the
time-averaged CO concentration obtained from the chemical evolution calculation
in the strongly dynamic simulation case. Dashed line: time-averaged CO concentration
corresponding to the ICE approximation, but calculating the
CO number densities of the atmospheric models
associated to each time step by using the same chemical evolution
code until reaching the ensuing equilibrium concentrations. Dotted line:
time-averaged CO concentration
corresponding to the ICE approximation, but calculating the
CO concentrations directly from the Saha chemical equilibrium equations.
A comparison of the dashed and dotted lines illustrates the reliability
of the chosen database for the chemical evolution calculations. In any case,
in order to be fully consistent with our comparisons, all ICE results in this paper refer to
`evolution until equilibrium' calculations.\label{fig2}}
\end{figure}

Fig. \ref{fig2} contrasts the time-averaged CO concentration obtained 
assuming ICE at each time step of the strongly dynamic simulation case 
with that resulting from the chemical evolution. 
Note that the ICE approximation leads to a significant overabundance of CO
in the outer layers of the atmosphere (i.e. above 700 km).
Thus, the CO number density can be relatively low in such outer layers, in spite
of the fact that the temporally averaged temperature 
of the Carlsson \& Stein (1997) simulations decreases outwards and does not show 
any chromospheric temperature rise. As expected from the previously reported results, 
the ICE approximation does a sufficiently 
good job concerning the modeling of the temporally-averaged CO spectrum
at the solar disk center ($\mu=1$). In contrast, the emergent CO spectrum computed 
close to the edge of the
solar disk (i.e. at $\mu=0.1$) shows significantly {\em stronger} CO-lines when the
ICE approximation is used, especially concerning strong CO lines
like the 3-2 R14 one, for which the line-core brightness temperature
is about 100 K {\em lower} than that obtained using the non-equilibrium CO number densities.

\begin{figure}
\epsscale{0.5}
\plotone{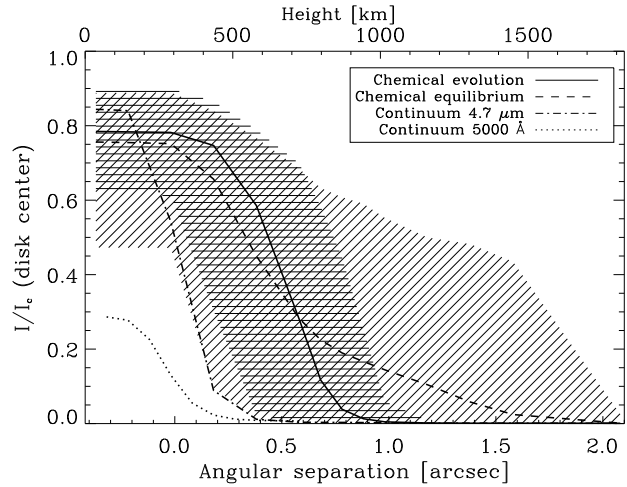}
\caption{The \emph{diagonal shading} indicates the relatively large
atmospheric region where the ICE approximation predicts off-limb emission
in the core of the 3-2 R14 CO-line, 
while the \emph{horizontal shading} shows that
corresponding to the chemical evolution calculation.
The figure also shows the calculated limb profiles for the continua at 4.7 $\mu$m and 
5000 \AA\ . The zero level of the angular separation 
scale refers to the 4.7 $\mu$m 
continuum edge calculated as the 50\% point in the intensity drop,
as normalized to the ensuing continuum intensity at disk center.  
At each angular distance, the
vertical extensions of the \emph{shadings} indicate the amplitude
of the temporal variation of the line-core CO-emission. The dashed and solid lines
give the corresponding temporally averaged off-limb emissions.
The upper horizontal axis gives the height in the atmosphere,
$z=0$ km being the location of continuum optical depth unity at 5000 \AA\ 
for a disk center observation (i.e., the usual depth scale in solar models).
Note that at low altitude locations, where the IR continuum dashed-dotted curve
lies above the (temporally-averaged) dashed and solid curves, the synthesized CO line 
turns into an absorption line for the ICE and non-equilibrium cases, respectively.   
\label{fig3}}
\end{figure}

Finally, we turn our attention to the modeling
issue of the off-limb CO emissions, which we have carried out
by solving the radiative transfer equation 
in spherical coordinates at each time step of the hydrodynamical simulation\footnote{This is
a suitable strategy for contrasting ICE and non-ICE results, as done also
in Fig. 1. It is however clear that
a truly realistic modeling of the available observations
should take into account the geometrical foreshortening effect resulting
from the fact that the solar chromosphere is a highly inhomogeneous 
medium in which the raypath goes through many different structures.}. 
The off-limb distances where (time-averaged) CO emission has been observed depend on 
the line (see Ayres 2002, for a summary of the available off-limb observational results): 
the off-limb emission extension of strong lines like the 3-2 R14 line 
lies between 0.55" and 0.7" above the 4.7 $\mu$m continuum limb, 
while weaker lines like the 7-6 R68 
line extend to $\sim$0.4". Fig. \ref{fig3} refers to the strongly dynamic case. It shows
that the atmospheric region where we can have
a significant off-limb emission is extremely large when the ICE approximation is used,
i.e. much larger than when the 
spectral synthesis is carried out using the non-equilibrium CO number densities.
The dashed and solid lines show the corresponding height variation 
of the temporally-averaged off-limb emission at the core of the
strong 3-2 R14 CO-line. 
They show that the non-equilibrium CO chemistry 
improves the agreement with the available off-limb observations. First, 
as seen in Fig. \ref{fig3}, the ICE approximation predicts 
that appreciable emission in the 3-2 R14 line should continue to relatively
large off-limb distances, while partial eclipse
measurements by Clark et al (1995) show a rapid disappearance of CO emission
at heights above 700 km, which is more in line 
with our chemical evolution calculations. Second, the representative 
off-limb emission extension where the normalized intensity falls to 50\%
of the on-disk value is larger when the non-equilibrium CO concentration
is used, which helps to improve the agreement with the
observations of translimb emission extensions,
although they still seem to be below the most recently observed values (see Ayres 2002).

\section{Concluding comment}
\label{conclusions} 

Our results indicate that the CO line radiation observed
close to the edge of the solar disk comes from 
atmospheric heights not greater than 700 km, approximately,
and that the gas in these regions of the low chromosphere 
must be much cooler than indicated by
the cool phases of the Carlsson \& Stein (1997) hydrodynamical simulations.
Lower temperatures will probably increase the relaxation times needed
to reach the molecular equilibrium concentrations. 
In a forthcoming publication we will show what happens when
the hydrodynamical simulations themselves are carried out taking the CO cooling
into account in a way consistent with the non-equilibrium
evolution of the molecular number densities. 

\section{Acknowledgements}

We thank Gene Avrett, Tom Ayres and Han Uitenbroek for useful remarks.  
This work has been supported by the Spanish Ministerio de Ciencia y Tecnolog\'\i a
through project AYA2001-1649, by the Norwegian Research Council through grant 146467/420
and by the European Commission via the Solar Magnetism Network.


\clearpage

\end{document}